\pdfoutput=1
\documentclass[%
 reprint,
nofootinbib,
 amsmath,amssymb,
 aps, prd,
floatfix,
showkeys,
]{revtex4-1}

\usepackage{graphicx}
\usepackage{dcolumn}
\usepackage{bm}
\usepackage{amsmath}
\usepackage{slashed}

\begin{document}

\preprint{}

\title{Scaling-correspondence mechanism}

\author{Jackie C.H. Liu}
 \email{chjliu@connect.ust.hk}%

\affiliation{%
Institute for Advanced Study, Hong Kong University of Science and Technology, Hong Kong
}%
\affiliation{%
Department of Physics, Hong Kong University of Science and Technology, Hong Kong
}%

\date{\today}

\begin{abstract}
In 1970s, Wilson shown the deep connection of renormalization and scaling of the effective Lagrangian.  Polchinski further proved that such connection implied renormalizability of perturbative field theory.  We develop the mechanism by an extension of scaling transformation of the effective Lagrangian - introducing a field-map between the quantum fields of the low-energy effective field theory (EFT) and its' scaling-corresponding (bare) model.  We study the mechanism through real scalar $\phi^4$ model.  We found if there is a self-interaction of order $N$ from the EFT's potential, there are (at most) $N-1$ scaling-corresponding models for the same symmetry group of the EFT.  We then study the potential application for cosmology by applying the mechanism to inflation paradigm.  We found the generated inflation model (from the real scalar $\phi^4$ EFT) is the power-law inflation model.  We briefly check the consistency of the generated inflation model against the inflation's typical observational constraints - spectral index ($n_{s}$), e-folds, and tensor/scalar ratio ($r$).

\end{abstract}

\maketitle

\section{\label{sec:level1}INTRODUCTION}

The understanding of physical scale and energy scale is crucial in particle physics, condensed matter research as well as cosmology.  For instance, conformal field theory (CFT) is an active research; it leads to AdS/CFT correspondence, which we can study symmetry restoration at high temperatures with the blackhole solutions \cite{Chai2020}.  The effect of beyond-standard-models may emerge beyond the energy scale of TeV \cite{Lykken,Astrid}  based on the measured Higgs mass and top quark mass, the perturbative analysis suggests that the electroweak Higgs vacuum might indeed not be stable; new degrees of freedom may appear below or at scales of $\mathcal{O}(10^{10})$ GeV, which changes the renormalization group evolution, and renders the potential stable.

Renormalization is one of key concepts in the development of quantum field theory \cite{Peskin}.  Wilson's work \cite{Wilson1971,Wilson1974} helps us understanding the renormalization in terms of relevant and irrelevant operators.  The deep connection of the renormalization and the scaling of effective Lagrangian was illustrated.  

In 1983, Polchinski advanced the idea from Wilson.  He shown that the connection found by Wilson can be made as the basis to prove the renormalizability of perturbative field theory \cite{Polchinski}.  Sean Carroll commented that "Indeed, there can be many possible 'true' theories - many 'ultraviolet completions' that would give you exactly the same low-energy effective theory!" \cite{Sean2013}.  Considering the perturbative EFT is the scaling-corresponding theory, i.e. scaled EFT, from the underlying bare field theory, we may ask if there is an extension of the scaling mechanism to yield a space of underlying field theories?  We propose such mechanism in this work.

Since the equations of motion (EoM) of the bare theory and the usual rescaled Lagrangian, $\mathcal{L}_{R}$, share the same form (also rescaled), we call two models are scaling-corresponding if there is such correspondence for their EoMs.  We introduce the field-map that maps the quantum field, $\phi$, from scaled EFT model to the quantum field, $\Phi$, of the underlying bare model.  For example, the field-map of scalar fields is defined by $\Phi(\phi)$.  The system described by the partition function of the underlying bare theory as, 
$\tilde{\mathcal{Z}}=\int \text{D$\Phi $} \, e^{i \tilde{S}\left(\Phi ,\tilde{g}\right)}$,
transforms as 
$\tilde{\mathcal{Z}}\to \int \text{D$\Phi(\phi) $} \, \text{D$\phi $}\, e^{i \tilde{S}\left(\Phi(\phi),\tilde{g}\right)}$, where $\tilde{S}$ and $\tilde{g}$ denote the related action and coupling(s) respectively.  The mechanism requires that the transformed system is equivalent to the system of the scaled EFT model when we impose the field-map and certain constraints.  We cover the construction of the mechanism and how it is related to Wilson's/Polchinski's approach for EFT in section 2.

In section 3, we develop a generic method to identify and solve the constraints.  We illustrate the mechanism through the real $\phi^4$ scalar field model.  We found that if there is a self-interaction of order $N$ in the EFT model, the solution (of order $N-1$ radical equation) consists of multiple models (up to $N-1$ models) for the underlying bare theory.
  In our illustrative example -  $\phi^4$ model, we use the two-loops renormalization from \cite{lezioni99}, and obtain the estimated cutoff of the EFT is at the scale of $\mathcal{O}(10^6)$ GeV.

In the last section, we further generalise the mechanism - introducing coupling-map $\tilde{g}(\Phi)$,  and apply to the field of cosmology - inflation paradigm.  Given the limited observational constraints of inflation cosmology \cite{Wang2014}, it can be useful if there is model-selection "guideline".  We demonstrate how the mechanism leads to the power-law model of slow-roll inflation from the $\phi^4$ EFT.
  We briefly verify the consistency of the generated inflation model by the observational constraints - the spectral index ($n_{s}$), e-folds, and tensor/scalar ratio ($r$) \cite{Leach2003, planck2018, Wang2014}.

\section{\label{sec:level1}The extension of scaling-correspondence mechanism}

In this section, we first briefly review the work developed by Wilson and Polchinski about the scaling of theories and renormalization as well as the analysis tool used by Polchinski.  We then introduce the mechanism as an extension of the scaling transformation, and explains how it may generate related underlying bare theories with some interesting properties.  Finally, we study the cutoff scale of the EFT model from the mechanism.

In the standard context of renormalization of perturbative quantum field theory to remove the infinities of loop calculation, we add the counter-terms to the renormalized terms, i.e. $\mathcal{L}_{R} + \mathcal{L}_{ct} $ \cite{Peskin}.  Wilson illustrated that the renormalization is indeed closely related to the scaling of the theory via the scaling transformation \cite{Wilson1971, Wilson1974}.  He used the irrelevant and relevant operators to analyse how the scaling and renormalization of EFT is connected.  In 1983, Polchinski further evaluated Wilson's idea in his paper "Renormalization and Effective Lagrangian" \cite{Polchinski}, and proved that the connection found by Wilson can be treated as the basis to prove the renormalizability of perturbative field theory.  We refer the reader to his work in detail of the proof \cite{Polchinski}.

\begin{figure}[htbp]
\includegraphics[width=180pt]{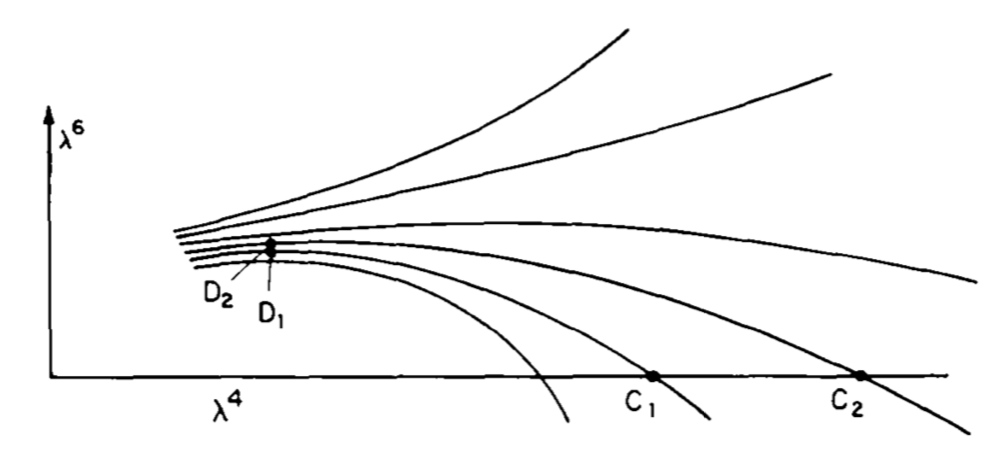}
\caption{\label{fig:conver} 
In $\lambda_4 -\lambda_6$ plane of the theory space of the example illustrated by Polchinski \cite{Polchinski}, the bare theories with different cutoffs at $C_1$ and $C_2$ moves forward to the convergence of trajectories at $D_1$ and $D_2$ as we move to the lower scale.}
\end{figure}

Polchinski illustrated by FIG. \ref{fig:conver} to show that, perturbative EFT is indeed in a "convergent" zone of the theory space, i.e. $\lambda_4 -\lambda_6$ plane, when performing scaling-transformation - scaling the bare theory at scale starting from $\Lambda_0$ to a much lower scale $\Lambda_R$.  The convergent zone means that the theories converge when we move up the cutoff scales.  In such example (4-dimensional scalar field theory), $\lambda_4$ is the coupling of the relevant operator (dimensionless); $\lambda_6$ is the coupling of the irrelevant operator (dimension of [$mass^{-2}$]); points $C_1$ and $C_2$ are initial points of the bare theories that subscript $1$ and $2$ denote different cutoff scales that the cutoff scales of point-2 is higher than the cutoff scales of point-1; points $D_1$ and $D_2$ are in convergent zone as we rescale the theories.  In convergent zone, relevant coupling $\lambda_4$ is the renormalized coupling, $\lambda^{R}_4$; in contrast, the irrelevant coupling is suppressed by $\mathcal{O}[\Lambda_{R}^2 / \Lambda_{0}^2]$.  

The EoMs associated with the theories for point $C_1$ and $C_2$ are the EoMs of bare theories, i.e. unscaled theories, while EoMs associated with the theories for point $D_1$ and $D_2$ are the EoMs of renormalized theories, i.e. scaled theories.  Both scaled and unscaled theories are in the same EoM form (with different coefficients only because of scaling).  We call two models are \textit{scaling-corresponding} if their EoMs are in the same form up to scaled coefficients.  Therefore, the bare theory, $\mathcal{L}_{B}$, and the perturbative renormalized EFT, $\mathcal{L}_{R}$, are scaling-corresponding.  In such scaling transformation, we have a specific \textit{field-map} (maps the quantum field of a system to the quantum field of another system): the usual field-rescaling, $\Phi = Z^{\frac{1}{2}}\; \phi$.

By introducing a general field-map, $\Phi(\phi)$, we extend the scaling transformation by postulating a mechanism that generates an underlying bare theory (or a set of underlying bare theories) $ \tilde{\mathcal{L}} $ of the quantum field $\Phi$ such that $ \tilde{\mathcal{L}} $ and the perturbative renormalized EFT of the quantum field $\phi$, $\mathcal{L}_{R}$, are scaling-corresponding.  The usual field-rescaling of renormalization procedure is the trivial case of the mechanism, e.g. $\Phi(\phi) = Z^{\frac{1}{2}}\; \phi$ with other scaled couplings.

We illustrate the mechanism by the scalar field theory for simplicity.  The system described by the partition function of the underlying bare theory is denoted as, 
\begin{equation}
\tilde{\mathcal{Z}}=\int \text{D$\Phi $} \, e^{i \tilde{S}\left(\Phi ,\tilde{g}\right)} \label{eq:Ztau},
\end{equation}
where $\tilde{S}$ and $\tilde{g}$ are the related action and coupling(s) respectively.  The field-map maps the scalar fields from $\phi$ to $\Phi$, i.e. $\Phi(\phi)$.  The system is then described by\footnote{We can add the terms $J \Phi$ and $J \phi$ to the Lagrangians of the actions of Eq. (\ref{eq:Ztau}) and Eq. (\ref{eq:Ztau2}) respectively to define the generating functional $\tilde{Z}[J]$.} 
\begin{equation}
\tilde{\mathcal{Z}}= \int \text{D$\Phi(\phi) $} \, \text{D$\phi $}\, e^{i \tilde{S}\left(\Phi(\phi), \tilde{g}\right)} \delta(\Omega) \label{eq:Ztau2},
\end{equation}
where $\Omega$ is the constraint to be identified; the field configurations are determined by both the field-map and $\phi$.  We require, given specific constraints and a field-map,  the system (\ref{eq:Ztau2}) is equivalent to the system of a perturbative renormalized EFT, which lies in the "\textit{finite-dimensional submanifold in the space of possible Lagrangians}", as referred by Polchinski \cite{Polchinski} (e.g. the theory lies in the convergent zone in FIG \ref{fig:conver}).  By equivalent, we mean the systems reproduce the same EFT in low energies.

In order to obtain the constraints and the field-map, we first obtain the EoM from a general form\footnote{We may postulate the potential by the sum of couplings with all combinations of field operators, e.g. $(\Phi_i)^{r} (\Phi_j)^{s}$.}
of $\tilde{\mathcal{L}}$, and expand the field-map around $\phi=v$ as
\begin{equation}
\Phi(\phi) = \Phi_0 + \Phi_{(1)} (\phi - v ) + \Phi_{(2)} \frac{(\phi - v)^2}{2} + ..., \label{eq:expansion}
\end{equation}
where $\Phi_{(r)} \equiv \Phi^{(r)}(v)$, then match such EoM to the EoM of the Lagrangian, $\mathcal{L}_{R} + \mathcal{L}_{ct}$, to obtain the constraints.  
Since the symmetry group of the EFT is related to the symmetry of the EoM, the constraints constructed from the couplings ($\tilde{g}_i$), $\Phi_0$ and $\Phi_{(r)}$, for each term of the EoM, are associated to the symmetry group's structure of the EFT.  
  The procedure is straight forward; we cover the generic method to solve the constraints and field-map with the examples in detail in the next section.


Considering the first order expansion for the field-map, $\Phi(\phi) \simeq \Phi_0 + \Phi_{(1)} (\phi - v )$, it is corresponded to the specific type of field redefinition, $\phi \to Z_\phi \phi + \phi_0$.  The kinetic term of the Lagrangian of the scalar field is trivially equivalent to the kinetic term of a rescaled field (by factor $Z_\phi $), and the new potential is equivalent to the shifted and rescaled field.  From \cite{1804}, the functional integral with the field redefinition is generally transformed to 
\begin{equation}
\mathcal{Z'}[J]= \int D\phi' \; e^{i \int \mathcal{L'}[\phi']+J\phi'},
\end{equation}
where $\mathcal{L'}$ and $\phi'$ are the transformed Lagrangian and scalar field respectively, and $\mathcal{L'}[\phi']=\mathcal{L}[\phi]$.  Note that the Jacobian of the functional measure, $| \frac{\delta F}{\delta \phi'} |$, is unity in dimensional regularization, where $F$ is the inverse map of field redefinition.  The green function is transformed but the S-matrix is not affected because the choice of interpolating field variable does not matter \cite{1804}.  Therefore, the field-mapped theory in the first order approximation reproduces the same S-matrix elements.

Under the context of EFT, we study the high order terms more carefully.  It is because expanding in the second order for the field-map, the kinetics term is transformed to the terms including interaction terms of new field and its derivative, e.g. $\phi^2  \partial{\phi}^2$.  From \cite{Peskin}, when one integrating out the high momentum modes, the functional integral can be expressed as
\begin{equation}
\mathcal{Z}= \int [D\phi]_{b \Lambda} \; e^{i \int \mathcal{L}_{\text{eff}}},
\end{equation}
where $\mathcal{L}_{\text{eff}} = \mathcal{L} + (\text{sum of connected diagrams})$.  The sum of the connected diagrams include the generated high dimensional terms such as the interaction terms of the field and its derivative because of Taylor expansion of momenta, and the Lagrangian is rescaled.  Therefore, the field-mapped theory would mix the generated high dimensional terms after field-map expansion with the connected diagrams when integrating out the high modes.  To be EFT in low energies, the high dimensional operators of EFT can be expressed as the following form \cite{1804},
\begin{equation}
\mathcal{L}_{\text{eff}} = \mathcal{L}_{\text{dim} \leq 4} + \frac{1}{\Lambda} \mathcal{C}^{(5)} \mathcal{O}^{(5)} + \frac{1}{\Lambda^2} \mathcal{C}^{(6)} \mathcal{O}^{(6)} + \text{...},\label{eq:EFTCond}
\end{equation}
where $ \mathcal{C}^{(r)}$ and $\mathcal{O}^{(r)}$ are the coefficients and the operators of high dimensional terms respectively.  Eq. (\ref{eq:EFTCond}) is the condition for the field-mapped theory to be equivalent to the original perturbative theory at low energies regime.  In the next section, after solving the field-maps, we check against such condition.

The previous two paragraphs explain the mechanism under the context of S-matrix elements, the functional integral, and the effect of integrating out high momentum modes.  The purpose is to show how the mechanism provides a consistent low-energies EFT.  Note that the EoMs-correspondence and Eq. (\ref{eq:EFTCond}) make the mechanism non-arbitrary because of such specific constraints.


Given the generated Lagrangian satisfying the condition - Eq. (\ref{eq:EFTCond}), one can use known EFT tools (e.g. removing redundant operators \cite{Brivio2019} and renormalization group equation (RGE) of SMEFT \cite{Jenkins2013}) to study the field-mapped theory\footnote{We expand the field-mapped Lagrangian which the KE term is in the canonical form.}.  
The high-dimensional operators of an EFT may lead to the non-trivial structure in the correlation functions \cite{Brivio2019}.  The dimension-six operators of SMEFT consist of total 59 independent operators, and the RGE is fairly complicated\cite{Jenkins2013}.  We consider such non-trivial-quantum-effects study in the future work.

Consider a more complicated scenario such as applying to Pati-Salam's extension of standard model \cite{Molinaro2018, PhysRevD.100.075009,PhysRevD.102.095025}, in which the symmetry group is
\begin{equation}
\text{SU(4)}\otimes \text{SU(2)}_L\otimes \text{SU(2)}_R \label{eq:PSSym}.
\end{equation}
The EoMs are more complicated, but the procedure is the same in principle - postulating a general form of $\tilde{\mathcal{L}}$ with all the possible terms and prefactor-coefficients as the couplings, expanding the field-maps (including scalar, gauge, and fermion fields), matching the EoMs, and then identifying the constraints.

The solution of the mechanism consisting of the constraints and the field-map, is generally not unique, because multiple solutions may exist.  Therefore, multiple bare theories may reproduce the same EFT model, and they usually have the similar structure as we illustrate in the next section by the $\phi^4$ model.

\begin{figure}[htbp]
\includegraphics[width=230pt]{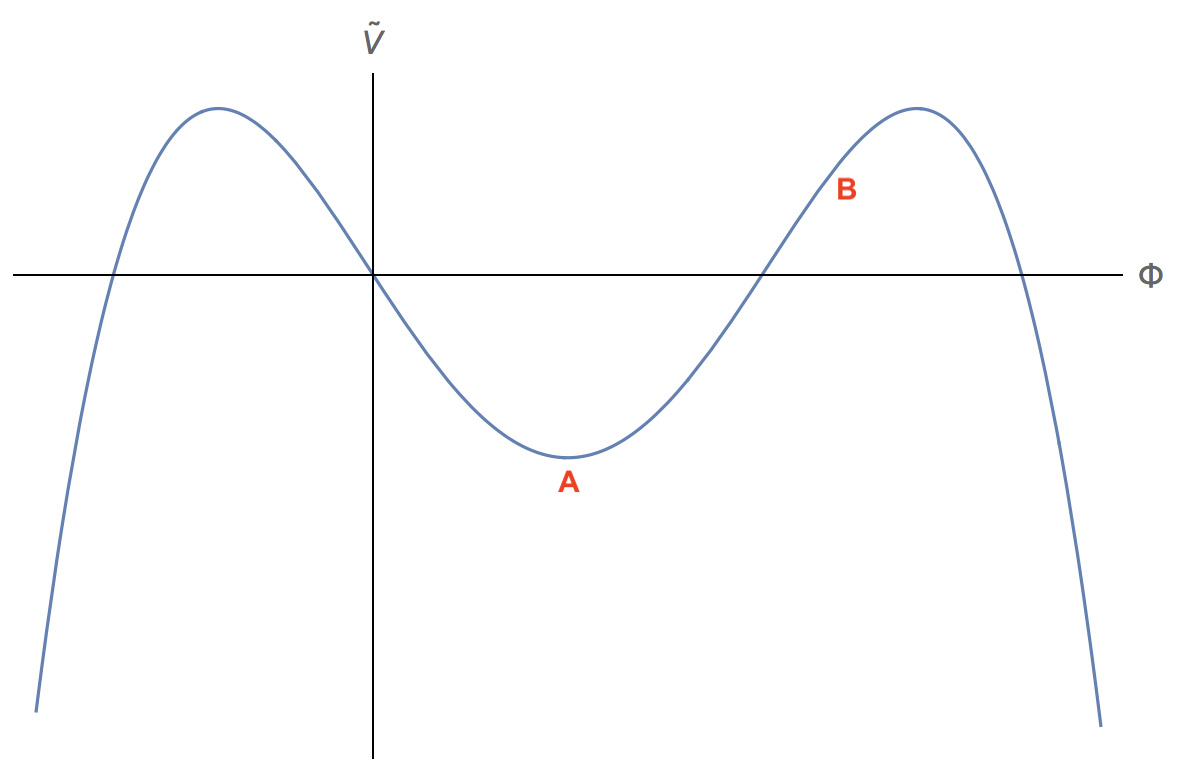}
\caption{\label{fig:VTau} 
Potential $\tilde{V}(\Phi)$ of an example of a bare theory with the inverted double-well potential.  In this example, the regime of validity of the mechanism is around the field value of point $A$ that the EFT model is reproduced, and becomes invalid at point $B$.}
\end{figure}

A reasonable cutoff scale of the EFT model is \textit{the energy scale of the validity of the solution} since the solution usually relies on the regime of the field $\Phi$ when we expand by Eq. (\ref{eq:expansion}) up to certain order.  For instance, in FIG. \ref{fig:VTau}, the valid regime of a bare theory with the inverted double-well potential is around the meta-stable point $A$ while the solution is invalid at point $B$\footnote{In the next section, we show how to justify the valid region by the suppression factor of the field-map expansion.}
; the field difference of point $A$ and $B$ indicates the energy scale threshold of the mechanism that the EFT model is reproduced.

In the Higgs mechanism \cite{Peskin}, the ground state of the potential causes the field to have a nonzero vacuum expected value, and as a result, below the electroweak-scale, the Higgs mechanism breaks the symmetry, and the field theory is described by the spontaneous symmetry breaking theory.  Similarly, we would speculate that, the ground state of the potential of $\tilde{\mathcal{L}}$ of the example in FIG. \ref{fig:VTau}, causes the field $\Phi$ to have a nonzero value, below the energy scale threshold, the mechanism of this work takes place such that the EFT emerges; above the energy scale threshold, the system is described by the related bare theory, $\tilde{\mathcal{L}}$.  However, the detail of the transition is not covered in this work.

\section{\label{sec:level1}Illustration for scalar field}

In this section, we explain how the mechanism works in detail by applying to the real scalar $\phi^4$ theory.

Here we demonstrate the generic method to obtain the constraints of the mechanism.  We assume the generic form of the bare theory of a real scalar field in Euclidean spacetime\footnote{For simplicity, we include only up to $\Phi ^4$.}
\begin{equation}
\tilde{\mathcal{L}}=\frac{1}{2} c_1 \partial _{\mu } \Phi  \partial ^{\mu } \Phi +\beta _0 \Phi+\frac{1}{2} m^2{}_0 \Phi ^2+\frac{\alpha _0 \Phi ^3}{6}+\frac{\lambda _0 \Phi ^4}{4!} \label{eq:LTau0},
\end{equation}
where $c_1$, $\beta _0$, $m^2{}_0$, $\alpha _0$, and $\lambda _0$ denote the coefficients. The related renormalized EFT is
 \begin{equation}
\mathcal{L}_{\text{ren}}=\frac{1}{2} Z_1 \partial _{\mu } \phi  \partial ^{\mu } \phi +\frac{1}{2} Z_{m^2} m_b^2   \phi ^2+\frac{Z_{\lambda } \lambda _b  \phi ^4}{4!} \label{eq:Lren},
\end{equation} 
where the renormalized $Z$-terms are two-loop renormalized \cite{lezioni99}.  After Taylor expanding the field map $\Phi(\phi)$ in the first order and matching term-by-terms for the EoMs of $\tilde{\mathcal{L}}$ and $\mathcal{L}_{\text{ren}}$, we obtain the constraints\footnote{We neglect the term $\phi^4$ of the EoM to match because it is associated with an irrelevant operator of $\mathcal{L}_{\text{ren}}$.}
\small  \begin{equation} \begin{split}
m^2{}_0 &=-\frac{2 \Phi _{\text{(1)}}^2 m_b^2 Z_{m^2}+\lambda _b Z_{\lambda } \left(\Phi _0-v \Phi _{\text{(1)}}\right){}^2}{2 \Phi _{\text{(1)}}^3} \\
\beta _0&=\frac{\left(\Phi _0-v \Phi _{\text{(1)}}\right) \left(6 \Phi _{\text{(1)}}^2 m_b^2 Z_{m^2}+\lambda _b Z_{\lambda } \left(\Phi _0-v \Phi _{\text{(1)}}\right){}^2\right)}{6 \Phi _{\text{(1)}}^3} \\
\alpha _0&=\frac{\lambda _b Z_{\lambda } \left(\Phi _0-v \Phi _{\text{(1)}}\right)}{\Phi _{\text{(1)}}^3},
\;\;  \lambda _0=-\frac{\lambda _b Z_{\lambda }}{\Phi _{\text{(1)}}^3}, 
\; \; c_1=-\frac{Z_1}{\Phi _{\text{(1)}}},
 \label{eq:cons}
\end{split} \end{equation}
and we can infer the \textit{field-map equation}
\begin{equation}
2 m^2{}_0 \Phi '(\phi )^3=\lambda _b Z_{\lambda } \left(\Phi (\phi )-v \Phi '(\phi )\right)^2-2 m_b^2 Z_{m^2} \Phi '(\phi )^2 \label{eq:radEqn}
\end{equation}
\normalsize
from the first constraint\footnote{One can also choose another constraint but we prefer the first one because it is related to a physical parameter - the mass-squared coefficient.}.  The similar method for relating a general potential to the effective field theory (around vacuum expected value) is found in the appendix of the literature \cite{Liu2018}, which considers the general Higgs potential parameterization to reproduce standard model effective field theory.

Interestingly, the field-map equation (\ref{eq:radEqn}) is a radical equation of order $N-1$, where $N$ is the highest power of $\phi$ of EFT, i.e. $N-1=3$ in case of $\phi^4$ scalar field.  Therefore, If there is a self-interaction of order $N$ in the EFT model, the solution consists of multiple models (up to $N-1$ models) for the underlying bare theory.  The bare theories are corresponded to the same EFT model's symmetry group, i.e. they reproduce the same symmetry of the EFT model.  We \textit{naively speculate} that the feature for having multiple bare models to the self-interaction EFT is rather general because the procedure to get the radical equation above is quite generic; however, we can't make the claim without a rigorous proof.

\begin{figure}[htbp]
\includegraphics[width=250pt]{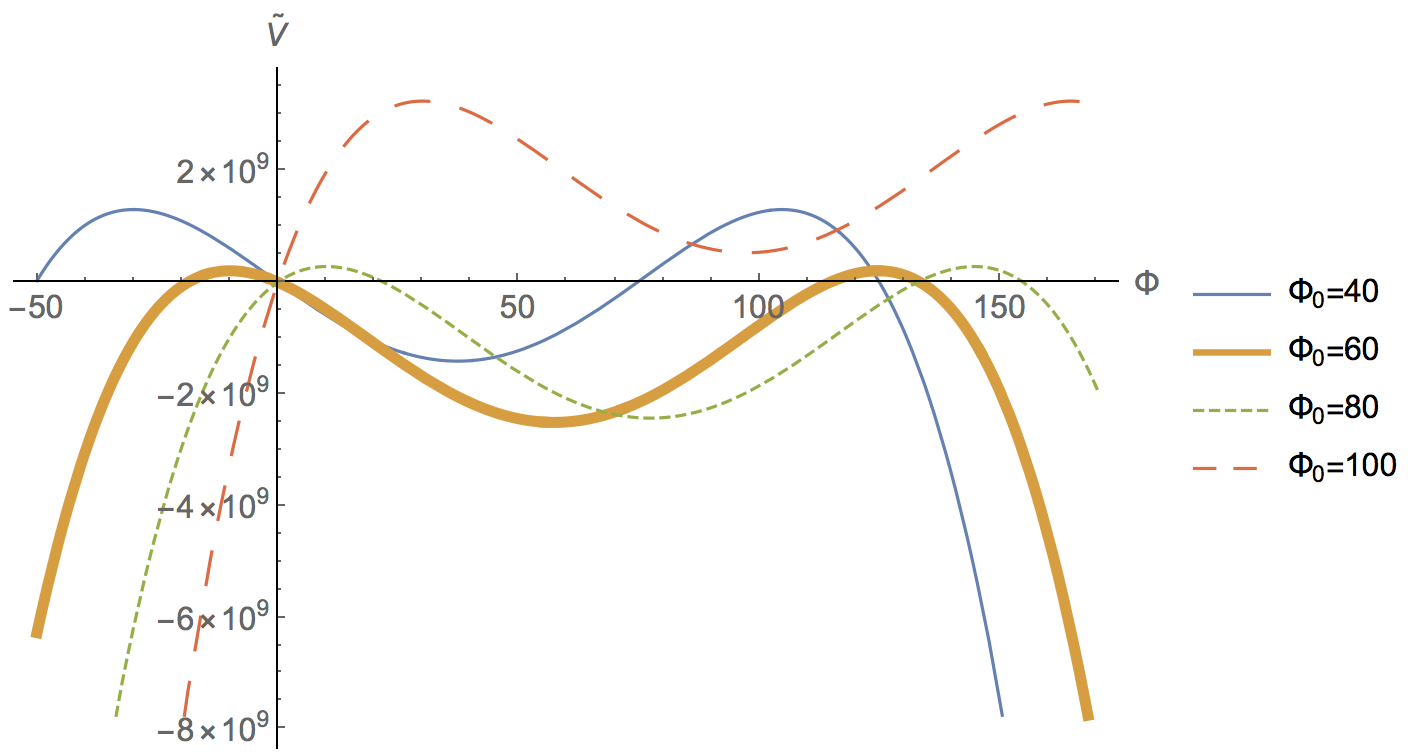}
\caption{\label{fig:VTau3} 
Potentials $\tilde{V}$ of the bare models with the inverted double-well potential generated by the mechanism for $\phi^4$ EFT (the mass of the scalar field of the EFT is set to 1).  We illustrate the potentials by four different $\Phi_0$ parameters.}
\end{figure}

To get the normalized kinetic term for $\tilde{\mathcal{L}}$, we set $c_1=1$.  The local minimum which we define as the vacuum expected value, VEV, for $\tilde{\mathcal{L}}$, is found
\begin{equation}
<\!\!\Phi\!\!> = v Z_1+\Phi _0 \label{eq:VEV}.
\end{equation}
FIG. \ref{fig:VTau3} is the illustrative plot for the potentials of $\tilde{\mathcal{L}}$ of different values of $\Phi_0$.  Note we assume a small finite $\epsilon$ parameter; the work Ref. \cite{Chai2020.2} treats $\epsilon$ a small finite parameter in the similar manner.

By re-defining $\Phi \to \Phi$ + VEV, and dropping constant, $Z_2$ symmetry is restored.  After substituting the leading order terms of the counter-terms from \cite{lezioni99}, we found the potential being inverted double-well and it's parameters are
\begin{equation} \begin{split}
\tilde{V}&=-\frac{12 m^2 \Phi ^2}{\epsilon }-\frac{216 \left(16 \pi ^2\right)^5 \Phi ^4 \epsilon ^2}{\lambda ^4}, \\ 
\tilde{\lambda }&=\frac{5184 \left(16 \pi ^2\right)^5 \epsilon ^2}{\lambda ^4}, 
\; \;  \tilde{m}=\frac{2 \sqrt{6} m}{\sqrt{\epsilon }}, \label{eq:VTauNPara}
\end{split} \end{equation}
where $m$ and $\lambda$ are the mass and quartic coupling of the scalar field of the EFT respectively.

The cubic roots of the field-map radical equation (\ref{eq:radEqn}), by $\epsilon$ expansion, lead to
\small \begin{align}
\Phi '(\phi )&=\frac{\lambda ^2}{3072 \pi ^4 \xi  \epsilon }-\frac{\lambda ^2 v^2}{256 \left(\pi ^2 \mu^2 \xi \right)}+\frac{24 \pi ^2 v \epsilon  \Phi (\phi )}{\mu^2}+O\left(\epsilon ^{3/2}\right)
 \\
\Phi '(\phi )&=\pm \frac{2 \sqrt{3} \pi  \sqrt{\epsilon } \left| \Phi (\phi )\right| }{\mu}-\frac{12 \epsilon  \left(\pi ^2 v \Phi (\phi )\right)}{\mu^2}+O\left(\epsilon ^{3/2}\right)  \label{eq:fieldmapEqns},
\end{align}
where $m\equiv i \mu$ and $m^2{}_0\equiv-\xi \, \tilde{m}^2$; the first solution of the field-map by leading terms and $\Phi_0$ are
\small \begin{equation} \begin{split}
\Phi (\phi )&=\frac{\lambda ^2 \mu^2 e^{-\frac{24 \pi ^2 v \epsilon  (v-\phi )}{\mu^2}}}{73728 \pi ^6 v \epsilon ^2} + C, \\
\Phi _0&=\frac{\lambda ^2 \left(\mu^2 (\xi -1)+12 \pi ^2 v^2 \epsilon \right)}{73728 \pi ^6 \xi  v \epsilon ^2}
 \label{eq:fieldmapSols1}
\end{split} \end{equation}
respectively, where $C$ is a constant; The second and third solutions of the field-map and $\Phi_0$ are
\begin{equation} \begin{split}
\Phi (\phi )&=-\frac{\lambda ^2 \mu^2 e^{\frac{2 \pi  \sqrt{\epsilon } (v-\phi ) \left(\pm\sqrt{3} \mu+6 \pi  v \sqrt{\epsilon }\right)}{\mu^2}}}{6144 \pi ^5 \epsilon ^{3/2} \left(\pm\sqrt{3} \mu+6 \pi  v \sqrt{\epsilon }\right)}, \\
\Phi _0&=-\frac{\lambda ^2 \mu^2}{6144 \pi ^5 \epsilon ^{3/2} \left(\pm\sqrt{3} \mu+6 \pi  v \sqrt{\epsilon }\right)}
 \label{eq:fieldmapSols2}
\end{split} \end{equation} \normalsize
respectively.  The suppression factor of the field-map solutions is $\mu/\!\sqrt{\epsilon}$ that consistently justifies the validity of the first-order expansion approximation in the low energies.  To check the first solution against the condition of EFT correspondence - Eq. (\ref{eq:EFTCond}), we found the coefficients of the high dimensional operators $\phi^5$,$\phi^6$,$\phi^7$, and $\phi^8$ are in order of $\mathcal{O}(\epsilon)$, $\mathcal{O}(\epsilon^2)$, $\mathcal{O}(\epsilon^3)$, and $\mathcal{O}(\epsilon^4)$ respectively\footnote{We expand the field-map up to second order in this work.}.  For the second and third solutions, the coefficients of the high dimensional operators $\phi^5$,$\phi^6$,$\phi^7$, and $\phi^8$ are in order of $\mathcal{O}(\epsilon^{1/2})$, $\mathcal{O}(\epsilon)$, $\mathcal{O}(\epsilon^{3/2})$, and $\mathcal{O}(\epsilon^2)$ respectively.  Therefore, the solution is valid because of the suppression factor $\epsilon$.

We naturally define the energy scale threshold (cutoff) of the EFT by the mass parameter $\tilde{m}$ of Eq. (\ref{eq:VTauNPara}).  The mechanism involving suppression factor is not uncommon; for instances, in \cite{Chen2017}, Boltzmann suppression is found that observing particles' mass much larger than the Hubble scale of the inflation is unlikely.  

We estimate the $\epsilon$ value to be $\sim3\times10^{-8}$ and the cutoff scale of $\phi^4$ EFT is in the order of $\mathcal{O}(10^6)$ GeV.  We use the running renormalisation group equation of $\lambda$ \cite{Peskin}, and coupling-matching-condition technique\footnote{By matching the coupling values of models of low energies and high energies at the cutoff.} \cite{Molinaro2018}, as well as the values of the Higgs boson's mass and vacuum expected value \cite{Khan2014}.  The smallness of $\epsilon$ is consistent with our assumption.  The seesaw mechanism of Pati-Salam model \cite{Molinaro2018} estimates the Pati-Salam breaking scale is at $\sim2000$ TeV.   And, the cutoff energy of an inverse seesaw model \cite{Nomura2019} to address the tiny-neutrinos issue is at around TeV scale.  Interestingly, the estimated order of magnitudes of the scales of those beyond-standard-models are fairly in-line with the cutoff scale calculated from our mechanism.

We can compare the VEVs by Eq (\ref{eq:VEV}).  The ratio of the VEVs of the first to the second (or third) solutions (Eq. (\ref{eq:fieldmapSols1}) and Eq. (\ref{eq:fieldmapSols2})) is $\sim\mathcal{O}(10^{-2}) / \! \sqrt{\epsilon}$.\footnote{We assume the parameter $\xi$ is in the order of unity.}  So, the VEVs of the solutions can have a \textit{hierarchy structure}.  With the $\epsilon$ estimation, the VEVs' ratio is $\sim\mathcal{O}(10^2)$.

Finally, we discuss briefly about the irrelevant operators in the perspective of our mechanism.  As mentioned in Polchinski's work \cite{Polchinski}, the effect of the irrelevant operators is present, but suppressed by $\mathcal{O}[\Lambda_{R}^2 / \Lambda_{0}^2]$.  To include the irrelevant operators, we can add the higher-dimensional operators with prefactors to Lagrangians of the Eq. (\ref{eq:LTau0}) and (\ref{eq:Lren}).  Then we follow the generic method in this section to obtain the constraints, and $N-1$ order radical equation for the field-map trivially.  If we can probe the related parameters of the high energy effect from the generated bare model by the observational data of early universe, in principle, we may examinate the low energy correction(s) by those irrelevant operators and validate the suppression factor $\mathcal{O}[\Lambda_{R}^2 / \Lambda_{0}^2]$.

\section{\label{sec:level1}Generating Inflaton model}

In this section, we further generalise the mechanism by introducing the \textit{coupling-maps} $\lambda_{i}(\Phi)$, and generate the inflaton model.  We demonstrate how the mechanism leads to the power-law inflaton model of the slow-roll inflation from the $\phi^4$ EFT.  For consistency checks, we compute the spectral index ($n_{s}$), e-folds, and tensor/scalar ratio ($r$), and compare them against the observational constraints in Ref. \cite{Leach2003, planck2018}.

We propose the general form of the underlying bare theory is
\small \begin{equation}
\frac{1}{2} c_1 \partial _{\mu } \Phi  \partial ^{\mu } \Phi + \lambda _1(\Phi ) \Phi +\frac{1}{2}  \lambda _2(\Phi ) \Phi ^2 +\frac{1}{3} \lambda _3(\Phi )\Phi ^3 +\frac{1}{4}  \lambda _4(\Phi )\Phi ^4  \label{eq:Linf0}.
\end{equation} \normalsize
As in the literature \cite{Chen2017}, the authors introduce the functions of the general parameterisation of the inflation-standard-model couplings, while we call the functions of the self-couplings the coupling-maps.  Following the same approach in last section, we match the EoMs, expand up to the first order for the coupling-maps around $\Phi_0$, and obtain the constraints.  We have the following coupling-maps differential equations:
\begin{widetext}
\footnotesize \begin{equation} \begin{split}
\lambda _1(\Phi )&=\frac{\left(\Phi -v \Phi '\right) \left(2 \left(6 m_b^2 Z_{m^2} \left(\Phi '\right)^2+\lambda _b Z_{\lambda } \left(\Phi -v \Phi '\right)^2\right)+3 \left(\Phi '\right)^3 \lambda _4'(\Phi ) \left(\Phi -v \Phi '\right)^3\right)}{12 \left(\Phi '\right)^3} \\
\lambda _2(\Phi )&=-\frac{2 m_b^2 Z_{m^2} \left(\Phi '\right)^2+\lambda _b Z_{\lambda } \left(\Phi -v \Phi '\right)^2}{2 \left(\Phi '\right)^3}-\lambda _1'(\Phi )-\lambda _4'(\Phi ) \left(\Phi -v \Phi '\right)^3 \\
\lambda _3(\Phi )&=\frac{\left(\Phi -v \Phi '\right) \left(\lambda _b Z_{\lambda }+3 \left(\Phi '\right)^3 \lambda _4'(\Phi ) \left(\Phi -v \Phi '\right)\right)}{2 \left(\Phi '\right)^3}-\frac{1}{2} \lambda _2'(\Phi ),
\;\;\;   \lambda _4(\Phi )=-\frac{\lambda _b Z_{\lambda }}{6 \left(\Phi '\right)^3}-\frac{1}{3} \lambda _3'(\Phi )+\lambda _4'(\Phi ) \left(v \Phi '-\Phi \right)
 \label{eq:couplingEqns},
\end{split} \end{equation} \normalsize
\end{widetext}
where $\Phi' = d\Phi / d\phi$.  The physical interpretation is that, the inflaton model consisting of coupling-maps becomes approximately the bare model of $\phi^4$ EFT at $\Phi\sim\Phi_0$.

\begin{figure}[htbp]
\includegraphics[width=250pt]{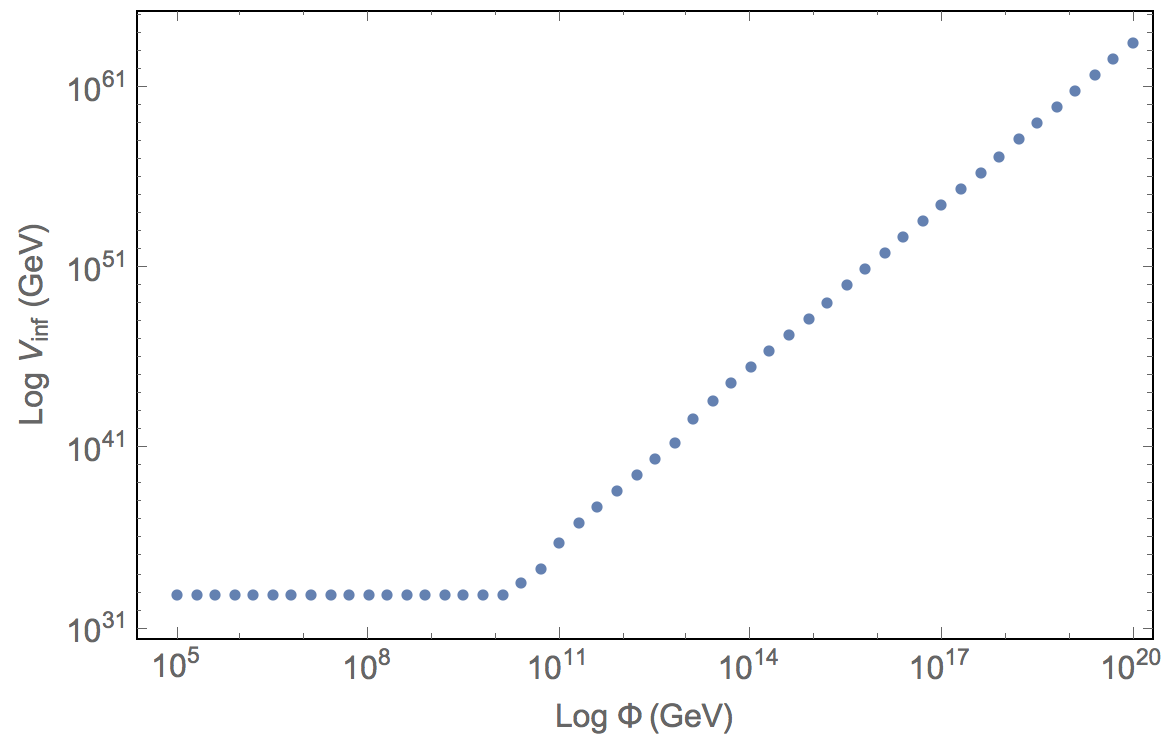}
\caption{\label{fig:VInf} 
Inflaton potential $V_{\text{inf}}$ generated by the mechanism and solved numerically.  In high-energies regime, the potential behaves as the power-law model.  We extrapolate the inflaton's potential in the power-law-form as $V_{\text{inf}} \propto \Phi ^{3.02}$}
\end{figure}

We set the boundary conditions for the coupling-maps by matching the model $\tilde{\mathcal{L}}$ in the previous section at $\Phi=\Phi_0$.  It implies the boundary conditions:
\small \begin{equation} \begin{split}
\lambda _1\left(\Phi _0\right)&=-\frac{\left(v Z_1+\Phi _0\right) \left(6 m_b^2 Z_{m^2} Z_1^2+Z_{\lambda } \lambda _b \left(v Z_1+\Phi _0\right){}^2\right)}{6 Z_1^3}, \\
 \lambda _2\left(\Phi _0\right)&=\frac{2 m_b^2 Z_{m^2} Z_1^2+Z_{\lambda } \lambda _b \left(v Z_1+\Phi _0\right){}^2}{2 Z_1^3}, \\
 \lambda _3\left(\Phi _0\right)&=-\frac{Z_{\lambda } \lambda _b \left(v Z_1+\Phi _0\right)}{2 Z_1^3},
\; \;  \lambda _4\left(\Phi _0\right)=\frac{Z_{\lambda } \lambda _b}{6 Z_1^3}.  
\end{split} \end{equation} \normalsize
In order to solve Eq. (\ref{eq:couplingEqns}), we need to specify the field-map from Eq. (\ref{eq:fieldmapSols1}) or Eq. (\ref{eq:fieldmapSols2}).  We choose the first solution as it is in higher energy scale to connect to the inflation scale, and we solve the system numerically\footnote{We assume the coupling parameter $\lambda=1$ and $\xi \sim \mathcal{O}(1)$ for simplicity.}.  The numerical plot is shown in FIG. \ref{fig:VInf}.

Although it is difficult to solve the system of Eq. (\ref{eq:couplingEqns}) analytically, interestingly, the inflaton's potential can be approximated by the power-law potential at high energies, 
\begin{equation}
V_{\text{inf}} \propto \Phi ^{\alpha} \label{eq:Vinf},
\end{equation}
as FIG. \ref{fig:VInf} shows.  We estimate the potential as $V_{\text{inf}} \propto \Phi ^{3.02}$.  Note that the power-law potential is the chaotic inflaton \cite{Baumann} (one of the large-field inflaton models).  Since Eq. (\ref{eq:couplingEqns}) is a parameterised differential system by the parameters from $Z$-terms of the EFT, we may expect the parameters of the $\phi^4$ EFT model are connected to such inflation model's parameter.  However, without the analytical solution, such relationship is difficult to justify.

By the extrapolated power-law potential from Eq. (\ref{eq:Vinf}), we compute the slow-roll parameters ($\epsilon_{\text{v}}$ and $\eta_{\text{v}}$) by formulas from \cite{Wang2014}, and found the inflation ends  $\Phi\sim\mathcal{O}(10^{18})$ GeV, about the Planck mass scale.  Given the $n_s$ formula from Ref. \cite{Pavluchenko2004}
\begin{equation}
n_s - 1 =2 \eta_{\text{v}}-6 \epsilon_{\text{v}} + \frac{1}{3} (44-18 c) \epsilon_{\text{v}}^2+(4 c-14) \eta_{\text{v}} \epsilon_{\text{v}}+\frac{2 \eta_{\text{v}}^2}{3} \label{eq:ns},
\end{equation}
where $c\simeq0.08145$, and $n_s = 0.97$ from \cite{planck2018}\footnote{We use the upper bound of the $n_s$.}, we calculated that the e-folds, the tensor/scalar ratio (r), and the energy scale of inflation are $\simeq83$, $0.14$, and $5\times10^{15}$ GeV respectively.  Comparing the typical expected e-folds ($>60$) \cite{Wang2014} and the inflation energy scale is 
\begin{equation}
3 < \frac{E_{\text{inf}}}{10^{15}\;\text{GeV}} < 29
\end{equation}
from Ref. \cite{Leach2003}, the e-folds and the energy scale of the generated inflaton model are fairly reasonable.   However there is tension for $r$, as the latest observational constraint \cite{planck2018} is $r< 0.10$.

The power-factor estimate of Eq. (\ref{eq:Vinf}), $\alpha\simeq3.0$, is within the bound from the observational constraint $\alpha\lesssim(3.5-4.5)$ from Ref. \cite{Pavluchenko2004}.
  Nevertheless, our work in this section is to demonstrate that the mechanism can be applied for the coupling-maps generalization; we expect more intensive study is needed in order to obtain more accurate inflation model.

\section{Discussion}

Understanding the scale of a physical system plays crucial role in different fields of Physics.  Inspired by Wilson's and Polchinski's work, we propose the mechanism by an extension of scale transformation for quantum field models.  We describe the mechanism by the example of real scalar $\phi^4$ effective field theory.  We found an interesting and generic feature that multiple bare theories ($N-1$) with similar structure of the potential are associated with the self-interaction of order $N$ from the EFT.  We further generalise the mechanism by introducing coupling functions, and apply to the inflation paradigm.  The generated inflaton is found to be power-law potential.

We expect the future work can improve the understanding of the transition from high-energies models to low-energies EFT, e.g. by the lattice method for field theory.  We may study the fermionic system with the gauge interaction of the symmetry group of the standard model by the mechanism and explore if there is hierarchy structure hinted in section 3.  Can we formulate an unification scheme for the families of the fermions?

We need to keep in mind that the tensor/scalar ratio deduced by generated inflaton in section 4 is in tension with the recent observation.  It is worth to pursue a more systemic analysis to improve the model and make some novel predictions that we can check against the observations such as cosmic non-Gaussianities.

\begin{acknowledgments}
JCHL would like to thank Professor Wang Yi of HKUST for the valuable comments and advices.
\end{acknowledgments}


\bibliography{export}

\begin{thebibliography}{25}%
\makeatletter
\providecommand \@ifxundefined [1]{%
 \@ifx{#1\undefined}
}%
\providecommand \@ifnum [1]{%
 \ifnum #1\expandafter \@firstoftwo
 \else \expandafter \@secondoftwo
 \fi
}%
\providecommand \@ifx [1]{%
 \ifx #1\expandafter \@firstoftwo
 \else \expandafter \@secondoftwo
 \fi
}%
\providecommand \natexlab [1]{#1}%
\providecommand \enquote  [1]{``#1''}%
\providecommand \bibnamefont  [1]{#1}%
\providecommand \bibfnamefont [1]{#1}%
\providecommand \citenamefont [1]{#1}%
\providecommand \href@noop [0]{\@secondoftwo}%
\providecommand \href [0]{\begingroup \@sanitize@url \@href}%
\providecommand \@href[1]{\@@startlink{#1}\@@href}%
\providecommand \@@href[1]{\endgroup#1\@@endlink}%
\providecommand \@sanitize@url [0]{\catcode `\\12\catcode `\$12\catcode
  `\&12\catcode `\#12\catcode `\^12\catcode `\_12\catcode `\%12\relax}%
\providecommand \@@startlink[1]{}%
\providecommand \@@endlink[0]{}%
\providecommand \url  [0]{\begingroup\@sanitize@url \@url }%
\providecommand \@url [1]{\endgroup\@href {#1}{\urlprefix }}%
\providecommand \urlprefix  [0]{URL }%
\providecommand \Eprint [0]{\href }%
\providecommand \doibase [0]{http://dx.doi.org/}%
\providecommand \selectlanguage [0]{\@gobble}%
\providecommand \bibinfo  [0]{\@secondoftwo}%
\providecommand \bibfield  [0]{\@secondoftwo}%
\providecommand \translation [1]{[#1]}%
\providecommand \BibitemOpen [0]{}%
\providecommand \bibitemStop [0]{}%
\providecommand \bibitemNoStop [0]{.\EOS\space}%
\providecommand \EOS [0]{\spacefactor3000\relax}%
\providecommand \BibitemShut  [1]{\csname bibitem#1\endcsname}%
\let\auto@bib@innerbib\@empty
\bibitem [{\citenamefont {Chai}\ \emph
  {et~al.}(2020{\natexlab{a}})\citenamefont {Chai}, \citenamefont {Chaudhuri},
  \citenamefont {Choi}, \citenamefont {Komargodski}, \citenamefont
  {Rabinovici},\ and\ \citenamefont {Smolkin}}]{Chai2020}%
  \BibitemOpen
  \bibfield  {author} {\bibinfo {author} {\bibfnamefont {N.}~\bibnamefont
  {Chai}}, \bibinfo {author} {\bibfnamefont {S.}~\bibnamefont {Chaudhuri}},
  \bibinfo {author} {\bibfnamefont {C.}~\bibnamefont {Choi}}, \bibinfo {author}
  {\bibfnamefont {Z.}~\bibnamefont {Komargodski}}, \bibinfo {author}
  {\bibfnamefont {E.}~\bibnamefont {Rabinovici}}, \ and\ \bibinfo {author}
  {\bibfnamefont {M.}~\bibnamefont {Smolkin}},\ }\href
  {https://arxiv.org/abs/2005.03676} {\bibfield  {journal} {\bibinfo  {journal}
  {Physical Review D}\ }\textbf {\bibinfo {volume} {102}},\ \bibinfo {pages}
  {065014} (\bibinfo {year} {2020}{\natexlab{a}})}\BibitemShut {NoStop}%
\bibitem [{\citenamefont {Lykken}(2010)}]{Lykken}%
  \BibitemOpen
  \bibfield  {author} {\bibinfo {author} {\bibfnamefont {J.~D.}\ \bibnamefont
  {Lykken}},\ }\href {\doibase 10.5170/CERN-2010-002.101} {\  (\bibinfo {year}
  {2010}),\ 10.5170/CERN-2010-002.101},\ \bibinfo {note}
  {arXiv:1005.1676}\BibitemShut {NoStop}%
\bibitem [{\citenamefont {Eichhorn}\ \emph {et~al.}(2015)\citenamefont
  {Eichhorn}, \citenamefont {Gies}, \citenamefont {Jaeckel}, \citenamefont
  {Plehn}, \citenamefont {Scherer},\ and\ \citenamefont
  {Sondenheimer}}]{Astrid}%
  \BibitemOpen
  \bibfield  {author} {\bibinfo {author} {\bibfnamefont {A.}~\bibnamefont
  {Eichhorn}}, \bibinfo {author} {\bibfnamefont {H.}~\bibnamefont {Gies}},
  \bibinfo {author} {\bibfnamefont {J.}~\bibnamefont {Jaeckel}}, \bibinfo
  {author} {\bibfnamefont {T.}~\bibnamefont {Plehn}}, \bibinfo {author}
  {\bibfnamefont {M.~M.}\ \bibnamefont {Scherer}}, \ and\ \bibinfo {author}
  {\bibfnamefont {R.}~\bibnamefont {Sondenheimer}},\ }\href {\doibase
  10.1007/JHEP04(2015)022} {\bibfield  {journal} {\bibinfo  {journal} {Journal
  of High Energy Physics}\ }\textbf {\bibinfo {volume} {2015}},\ \bibinfo
  {pages} {22} (\bibinfo {year} {2015})}\BibitemShut {NoStop}%
\bibitem [{\citenamefont {Peskin}\ and\ \citenamefont
  {Schroeder}(1995)}]{Peskin}%
  \BibitemOpen
  \bibfield  {author} {\bibinfo {author} {\bibfnamefont {M.~E.}\ \bibnamefont
  {Peskin}}\ and\ \bibinfo {author} {\bibfnamefont {D.~V.}\ \bibnamefont
  {Schroeder}},\ }\href {https://books.google.com.hk/books?id=EVeNNcslvX0C}
  {\emph {\bibinfo {title} {An Introduction To Quantum Field Theory}}}\
  (\bibinfo  {publisher} {Avalon Publishing},\ \bibinfo {year}
  {1995})\BibitemShut {NoStop}%
\bibitem [{\citenamefont {Wilson}(1971)}]{Wilson1971}%
  \BibitemOpen
  \bibfield  {author} {\bibinfo {author} {\bibfnamefont {K.~G.}\ \bibnamefont
  {Wilson}},\ }\href {\doibase 10.1103/PhysRevB.4.3174} {\bibfield  {journal}
  {\bibinfo  {journal} {Phys.Rev.B}\ }\textbf {\bibinfo {volume} {4}},\
  \bibinfo {pages} {3174} (\bibinfo {year} {1971})}\BibitemShut {NoStop}%
\bibitem [{\citenamefont {Wilson}\ and\ \citenamefont
  {Kogut}(1974)}]{Wilson1974}%
  \BibitemOpen
  \bibfield  {author} {\bibinfo {author} {\bibfnamefont {K.~G.}\ \bibnamefont
  {Wilson}}\ and\ \bibinfo {author} {\bibfnamefont {J.}~\bibnamefont {Kogut}},\
  }\href {\doibase https://doi.org/10.1016/0370-1573(74)90023-4} {\bibfield
  {journal} {\bibinfo  {journal} {Physics Reports}\ }\textbf {\bibinfo {volume}
  {12}},\ \bibinfo {pages} {75} (\bibinfo {year} {1974})}\BibitemShut {NoStop}%
\bibitem [{\citenamefont {Polchinski}(1984)}]{Polchinski}%
  \BibitemOpen
  \bibfield  {author} {\bibinfo {author} {\bibfnamefont {J.}~\bibnamefont
  {Polchinski}},\ }\href {\doibase
  https://doi.org/10.1016/0550-3213(84)90287-6} {\bibfield  {journal} {\bibinfo
   {journal} {Nuclear Physics B}\ }\textbf {\bibinfo {volume} {231}},\ \bibinfo
  {pages} {269} (\bibinfo {year} {1984})}\BibitemShut {NoStop}%
\bibitem [{\citenamefont {Carroll}(2013)}]{Sean2013}%
  \BibitemOpen
  \bibfield  {author} {\bibinfo {author} {\bibfnamefont {S.}~\bibnamefont
  {Carroll}},\ }\href@noop {} {\emph {\bibinfo {title} {How Quantum Field
  Theory Becomes Effective}}}\ (\bibinfo {year} {2013})\ \bibinfo {note}
  {www.preposterousuniverse.com}\BibitemShut {NoStop}%
\bibitem [{\citenamefont {Casalbuoni}(2005)}]{lezioni99}%
  \BibitemOpen
  \bibfield  {author} {\bibinfo {author} {\bibfnamefont {R.}~\bibnamefont
  {Casalbuoni}},\ }\href@noop {} {\emph {\bibinfo {title} {Advanced Quantum
  Field Theory}}}\ (\bibinfo {year} {2004/2005})\ \bibinfo {note} {lezioni date
  all universita di Firenze, Dipartimento di Fisica, Firence Italy}\BibitemShut
  {NoStop}%
\bibitem [{\citenamefont {Wang}(2014)}]{Wang2014}%
  \BibitemOpen
  \bibfield  {author} {\bibinfo {author} {\bibfnamefont {Y.}~\bibnamefont
  {Wang}},\ }\href {\doibase 10.1088/0253-6102/62/1/19} {\bibfield  {journal}
  {\bibinfo  {journal} {Communications in Theoretical Physics}\ }\textbf
  {\bibinfo {volume} {62}} (\bibinfo {year} {2014}),\
  10.1088/0253-6102/62/1/19}\BibitemShut {NoStop}%
\bibitem [{\citenamefont {Leach}\ and\ \citenamefont
  {Liddle}(2003)}]{Leach2003}%
  \BibitemOpen
  \bibfield  {author} {\bibinfo {author} {\bibfnamefont {S.~M.}\ \bibnamefont
  {Leach}}\ and\ \bibinfo {author} {\bibfnamefont {A.~R.}\ \bibnamefont
  {Liddle}},\ }\href {\doibase 10.1046/j.1365-8711.2003.06445.x} {\bibfield
  {journal} {\bibinfo  {journal} {Monthly Notices of the Royal Astronomical
  Society}\ }\textbf {\bibinfo {volume} {341}},\ \bibinfo {pages} {1151}
  (\bibinfo {year} {2003})}\BibitemShut {NoStop}%
\bibitem [{\citenamefont {Akrami}\ \emph {et~al.}(2020)\citenamefont {Akrami}
  \emph {et~al.}}]{planck2018}%
  \BibitemOpen
  \bibfield  {author} {\bibinfo {author} {\bibfnamefont {Y.}~\bibnamefont
  {Akrami}} \emph {et~al.},\ }\href
  {https://doi.org/10.1051/0004-6361/201833887} {\bibfield  {journal} {\bibinfo
   {journal} {A \& A}\ }\textbf {\bibinfo {volume} {641}} (\bibinfo {year}
  {2020})}\BibitemShut {NoStop}%
\bibitem [{\citenamefont {Manohar}(2018)}]{1804}%
  \BibitemOpen
  \bibfield  {author} {\bibinfo {author} {\bibfnamefont {A.~V.}\ \bibnamefont
  {Manohar}},\ }\href@noop {} {\enquote {\bibinfo {title} {Introduction to
  effective field theories},}\ } (\bibinfo {year} {2018}),\ \Eprint
  {http://arxiv.org/abs/1804.05863} {arXiv:1804.05863 [hep-ph]} \BibitemShut
  {NoStop}%
\bibitem [{\citenamefont {Brivio}\ and\ \citenamefont
  {Trott}(2019)}]{Brivio2019}%
  \BibitemOpen
  \bibfield  {author} {\bibinfo {author} {\bibfnamefont {I.}~\bibnamefont
  {Brivio}}\ and\ \bibinfo {author} {\bibfnamefont {M.}~\bibnamefont {Trott}},\
  }\href {\doibase 10.1016/j.physrep.2018.11.002} {\bibfield  {journal}
  {\bibinfo  {journal} {Physics Reports}\ }\textbf {\bibinfo {volume} {793}},\
  \bibinfo {pages} {1–98} (\bibinfo {year} {2019})}\BibitemShut {NoStop}%
\bibitem [{\citenamefont {Jenkins}\ \emph {et~al.}(2013)\citenamefont
  {Jenkins}, \citenamefont {Manohar},\ and\ \citenamefont
  {Trott}}]{Jenkins2013}%
  \BibitemOpen
  \bibfield  {author} {\bibinfo {author} {\bibfnamefont {E.~E.}\ \bibnamefont
  {Jenkins}}, \bibinfo {author} {\bibfnamefont {A.~V.}\ \bibnamefont
  {Manohar}}, \ and\ \bibinfo {author} {\bibfnamefont {M.}~\bibnamefont
  {Trott}},\ }\href {\doibase 10.1007/jhep10(2013)087} {\bibfield  {journal}
  {\bibinfo  {journal} {Journal of High Energy Physics}\ }\textbf {\bibinfo
  {volume} {2013}} (\bibinfo {year} {2013}),\
  10.1007/jhep10(2013)087}\BibitemShut {NoStop}%
\bibitem [{\citenamefont {Molinaro}\ \emph {et~al.}(2018)\citenamefont
  {Molinaro}, \citenamefont {Sannino},\ and\ \citenamefont
  {Wang}}]{Molinaro2018}%
  \BibitemOpen
  \bibfield  {author} {\bibinfo {author} {\bibfnamefont {E.}~\bibnamefont
  {Molinaro}}, \bibinfo {author} {\bibfnamefont {F.}~\bibnamefont {Sannino}}, \
  and\ \bibinfo {author} {\bibfnamefont {Z.~W.}\ \bibnamefont {Wang}},\ }\href
  {\doibase 10.1103/PhysRevD.98.115007} {\bibfield  {journal} {\bibinfo
  {journal} {Phys.Rev.D}\ }\textbf {\bibinfo {volume} {98}},\ \bibinfo {pages}
  {115007} (\bibinfo {year} {2018})}\BibitemShut {NoStop}%
\bibitem [{\citenamefont {Sannino}\ \emph {et~al.}(2019)\citenamefont
  {Sannino}, \citenamefont {Smirnov},\ and\ \citenamefont
  {Wang}}]{PhysRevD.100.075009}%
  \BibitemOpen
  \bibfield  {author} {\bibinfo {author} {\bibfnamefont {F.}~\bibnamefont
  {Sannino}}, \bibinfo {author} {\bibfnamefont {J.}~\bibnamefont {Smirnov}}, \
  and\ \bibinfo {author} {\bibfnamefont {Z.~W.}\ \bibnamefont {Wang}},\ }\href
  {\doibase 10.1103/PhysRevD.100.075009} {\bibfield  {journal} {\bibinfo
  {journal} {Phys. Rev. D}\ }\textbf {\bibinfo {volume} {100}},\ \bibinfo
  {pages} {075009} (\bibinfo {year} {2019})}\BibitemShut {NoStop}%
\bibitem [{\citenamefont {Huang}\ \emph {et~al.}(2020)\citenamefont {Huang},
  \citenamefont {Sannino},\ and\ \citenamefont {Wang}}]{PhysRevD.102.095025}%
  \BibitemOpen
  \bibfield  {author} {\bibinfo {author} {\bibfnamefont {W.~C.}\ \bibnamefont
  {Huang}}, \bibinfo {author} {\bibfnamefont {F.}~\bibnamefont {Sannino}}, \
  and\ \bibinfo {author} {\bibfnamefont {Z.~W.}\ \bibnamefont {Wang}},\ }\href
  {\doibase 10.1103/PhysRevD.102.095025} {\bibfield  {journal} {\bibinfo
  {journal} {Phys. Rev. D}\ }\textbf {\bibinfo {volume} {102}},\ \bibinfo
  {pages} {095025} (\bibinfo {year} {2020})}\BibitemShut {NoStop}%
\bibitem [{\citenamefont {Liu}\ \emph {et~al.}(2018)\citenamefont {Liu},
  \citenamefont {Lyu}, \citenamefont {Ren},\ and\ \citenamefont
  {Zhu}}]{Liu2018}%
  \BibitemOpen
  \bibfield  {author} {\bibinfo {author} {\bibfnamefont {T.}~\bibnamefont
  {Liu}}, \bibinfo {author} {\bibfnamefont {K.-F.}\ \bibnamefont {Lyu}},
  \bibinfo {author} {\bibfnamefont {J.}~\bibnamefont {Ren}}, \ and\ \bibinfo
  {author} {\bibfnamefont {H.~X.}\ \bibnamefont {Zhu}},\ }\href {\doibase
  10.1103/PhysRevD.98.093004} {\bibfield  {journal} {\bibinfo  {journal}
  {Phys.Rev.D}\ }\textbf {\bibinfo {volume} {98}},\ \bibinfo {pages} {093004}
  (\bibinfo {year} {2018})}\BibitemShut {NoStop}%
\bibitem [{\citenamefont {Chai}\ \emph
  {et~al.}(2020{\natexlab{b}})\citenamefont {Chai}, \citenamefont {Chaudhuri},
  \citenamefont {Choi}, \citenamefont {Komargodski}, \citenamefont
  {Rabinovici},\ and\ \citenamefont {Smolkin}}]{Chai2020.2}%
  \BibitemOpen
  \bibfield  {author} {\bibinfo {author} {\bibfnamefont {N.}~\bibnamefont
  {Chai}}, \bibinfo {author} {\bibfnamefont {S.}~\bibnamefont {Chaudhuri}},
  \bibinfo {author} {\bibfnamefont {C.}~\bibnamefont {Choi}}, \bibinfo {author}
  {\bibfnamefont {Z.}~\bibnamefont {Komargodski}}, \bibinfo {author}
  {\bibfnamefont {E.}~\bibnamefont {Rabinovici}}, \ and\ \bibinfo {author}
  {\bibfnamefont {M.}~\bibnamefont {Smolkin}},\ }\href {\doibase
  10.1103/PhysRevLett.125.131603} {\bibfield  {journal} {\bibinfo  {journal}
  {Phys.Rev.Lett.}\ }\textbf {\bibinfo {volume} {125}},\ \bibinfo {pages}
  {131603} (\bibinfo {year} {2020}{\natexlab{b}})}\BibitemShut {NoStop}%
\bibitem [{\citenamefont {Chen}\ \emph {et~al.}(2017)\citenamefont {Chen},
  \citenamefont {Wang},\ and\ \citenamefont {Xianyu}}]{Chen2017}%
  \BibitemOpen
  \bibfield  {author} {\bibinfo {author} {\bibfnamefont {X.}~\bibnamefont
  {Chen}}, \bibinfo {author} {\bibfnamefont {Y.}~\bibnamefont {Wang}}, \ and\
  \bibinfo {author} {\bibfnamefont {Z.-Z.}\ \bibnamefont {Xianyu}},\ }\href
  {\doibase 10.1103/PhysRevLett.118.261302} {\bibfield  {journal} {\bibinfo
  {journal} {Phys.Rev.Lett.}\ }\textbf {\bibinfo {volume} {118}},\ \bibinfo
  {pages} {261302} (\bibinfo {year} {2017})}\BibitemShut {NoStop}%
\bibitem [{\citenamefont {Khan}\ and\ \citenamefont
  {Rakshit}(2014)}]{Khan2014}%
  \BibitemOpen
  \bibfield  {author} {\bibinfo {author} {\bibfnamefont {N.}~\bibnamefont
  {Khan}}\ and\ \bibinfo {author} {\bibfnamefont {S.}~\bibnamefont {Rakshit}},\
  }\href {\doibase 10.1103/PhysRevD.90.113008} {\bibfield  {journal} {\bibinfo
  {journal} {Phys.Rev.D}\ }\textbf {\bibinfo {volume} {90}},\ \bibinfo {pages}
  {113008} (\bibinfo {year} {2014})}\BibitemShut {NoStop}%
\bibitem [{\citenamefont {Nomura}\ and\ \citenamefont
  {Okada}(2019)}]{Nomura2019}%
  \BibitemOpen
  \bibfield  {author} {\bibinfo {author} {\bibfnamefont {T.}~\bibnamefont
  {Nomura}}\ and\ \bibinfo {author} {\bibfnamefont {H.}~\bibnamefont {Okada}},\
  }\href {\doibase 10.1103/PhysRevD.99.055027} {\bibfield  {journal} {\bibinfo
  {journal} {Phys.Rev.D}\ }\textbf {\bibinfo {volume} {99}},\ \bibinfo {pages}
  {055027} (\bibinfo {year} {2019})}\BibitemShut {NoStop}%
\bibitem [{\citenamefont {Baumann}(2009)}]{Baumann}%
  \BibitemOpen
  \bibfield  {author} {\bibinfo {author} {\bibfnamefont {D.}~\bibnamefont
  {Baumann}},\ }\href@noop {} {\emph {\bibinfo {title} {TASI Lectures on
  Inflation}}}\ (\bibinfo {year} {2009})\ \bibinfo {note}
  {arXiv:0907.5424v2}\BibitemShut {NoStop}%
\bibitem [{\citenamefont {Pavluchenko}(2004)}]{Pavluchenko2004}%
  \BibitemOpen
  \bibfield  {author} {\bibinfo {author} {\bibfnamefont {S.~A.}\ \bibnamefont
  {Pavluchenko}},\ }\href {\doibase 10.1103/PhysRevD.69.021301} {\bibfield
  {journal} {\bibinfo  {journal} {Phys.Rev.D}\ }\textbf {\bibinfo {volume}
  {69}},\ \bibinfo {pages} {021301(R)} (\bibinfo {year} {2004})}\BibitemShut
  {NoStop}%
\end{thebibliography}%

\end{document}